\documentclass[conference]{IEEEtran}
\IEEEoverridecommandlockouts

\usepackage{cite}
\usepackage{amsmath,amssymb,amsfonts}
\usepackage{algorithmic}
\usepackage{graphicx}
\usepackage{textcomp}
\usepackage{xcolor}
\def\BibTeX{{\rm B\kern-.05em{\sc i\kern-.025em b}\kern-.08em
    T\kern-.1667em\lower.7ex\hbox{E}\kern-.125emX}}

\usepackage{hyperref}
\usepackage{url}
\usepackage{booktabs}
\usepackage{threeparttable}

\newcommand{\nummalware}{133 }
\newcommand{\numperformance}{11 }

\begin{document}

\title{MaLAware: Automating the Comprehension of Malicious Software Behaviours using Large Language Models (LLMs)\\

}

\author{\IEEEauthorblockN{Bikash Saha, Nanda Rani, Sandeep Kumar Shukla}
\IEEEauthorblockA{\textit{Department of Computer Science \& Engineering} \\
\textit{Indian Institute of Technology Kanpur}\\
Kanpur, India \\
\{bikash,nandarani,sandeeps\}@cse.iitk.ac.in}
}

\maketitle

\begin{abstract}

Current malware (malicious software) analysis tools focus on detection and family classification but fail to provide clear and actionable narrative insights into the malignant activity of the malware. Therefore, there is a need for a tool that translates raw malware data into human-readable descriptions. Developing such a tool accelerates incident response, reduces malware analysts' cognitive load, and enables individuals having limited technical expertise to understand malicious software behaviour. With this objective, we present MaLAware, which automatically summarizes the full spectrum of malicious activity of malware executables. MaLAware processes Cuckoo Sandbox-generated reports using large language models (LLMs) to correlate malignant activities and generate concise summaries explaining malware behaviour. We evaluate the tool's performance on five open-source LLMs. The evaluation uses the human-written malware behaviour description dataset as ground truth. The model's performance is measured using 11 extensive performance metrics, which boosts the confidence of MaLAware's effectiveness. The current version of the tool, i.e., MaLAware, supports Qwen2.5-7B, Llama2-7B, Llama3.1-8B, Mistral-7B, and Falcon-7B, along with the quantization feature for resource-constrained environments. MaLAware lays a foundation for future research in malware behavior explanation, and its extensive evaluation demonstrates LLMs' ability to narrate malware behavior in an actionable and comprehensive manner.

\end{abstract}


\section{Introduction}
\label{sec:MaLAwareIntroduction}

The problem of malicious software is rising exponentially worldwide. 
While malware detection and family classification have been extensively studied~\cite{or2019dynamic,aslan2020comprehensive,gopinath2023comprehensive,rani2022survey}, existing approaches often fail to capture the nuanced behaviors of malicious software and lack a structured, interpretable presentation of these insights for practical use.
Analysts face the burden of manually analyzing sandbox reports, which is time-consuming and mentally taxing. This cumbersome process delays incident response and hinders the implementation of tailored remediation. We present MaLAware, a tool that effectively comprehends and summarizes the  full spectrum of malware activities into human-readable text.

Consider a scenario where a cybersecurity analyst identifies new malware infiltrating a company's network. Traditional tools classify it as malicious and assign a known family but provide only technical jargon and raw data. The analyst is required to manually analyze intricate sandbox reports to comprehend the full range of malware behavior~\cite{saha2023malxcap}. This approach is cognitively demanding, leading to delays in response and increased organizational risk. Meanwhile, management demands clear explanations to inform stakeholders and guide actions.

The proposed MaLAware tool simplifies this process. It quickly generates clear, human-readable summaries of malware activities by parsing sandbox reports and leveraging large language models (LLMs). 
This reduces the analyst's cognitive load and accelerates incident response. For instance, the analyst can promptly report that the malware is exfiltrating customer data via encrypted channels and recommend network isolation and enhanced security measures.

MaLAware's concise summaries enable team members, including executives and legal advisors, to grasp the severity of threats without needing extensive technical expertise. This fosters faster decision-making and coordinated efforts to mitigate risks. By automating complex behavior explanations, MaLAware bridges a critical gap in existing tools, improving efficiency for cybersecurity teams and strengthening organizational security.

Our proposed MaLAware\footnote{Available at:\url{https://github.com/bikasaha/MaLAware}} is a tool
that generates structured, human-readable descriptions of malicious activities.
This tool enhances understanding of malware activities by combining Cuckoo Sandbox~\cite{cuckooSandbox} analysis with LLMs to interpret and summarize malware behaviors in a human-readable textual format. 

MaLAware processes cuckoo sandbox reports (in JSON format) by filtering relevant sections. These sections are fed into the LLM, which analyzes and correlates the malware's various events and actions during execution to generate a comprehensive textual summary. Finally, post-processing scripts refine the LLM’s output to produce well-structured and comprehensive summaries. 
To evaluate the model efficiency, we leverage our expertise in malware and create a dataset
detailing the malicious activities of \nummalware malware samples. 
We evaluate MaLAware using five LLMs—Qwen2.5-7B\cite{hui2024qwen2}, Llama2-7B\cite{touvron2023llama}, Llama3.1-8B\cite{dubey2024llama}, Mistral-7B\cite{jiang2023mistral}, and Falcon-7B~\cite{almazrouei2023falcon}—and assess performance across \numperformance diverse metrics for a comprehensive comparison. The current version of MaLAware supports all five LLMs and includes a quantization feature to let the tool run in resource-constrained environments.

We publicly release the tool to support the research community. The tool, along with detailed usage instructions and sample dataset, is available at the following link: \url{https://doi.org/10.5281/zenodo.14809826}.

Section~\ref{sec:MaLAwarePotentialApplication} presents the potential applications and use cases of our tool. In section~\ref{sec:MaLAware}, we discuss the workflow of MaLAware including installation and execution details. We discuss evaluation including dataset in Section~\ref{sec:MaLAwareEvaluation}. The limitations in the current version and our future work are discussed in Section~\ref{sec:MaLAwareLimitationFutureWork}. Further, Section~\ref{sec:MaLAwareRelatedWork} discusses the current literature and we concludes the contribution in Section~\ref{sec:MaLAwareConclusion}.

\section{Potential Application and Use-cases}
\label{sec:MaLAwarePotentialApplication}

Understanding and comprehending malware behavior through MaLAware can significantly enhance incident response by enabling quicker threat mitigation and improving threat intelligence by providing deeper insights into adversarial tactics. Its integration into cybersecurity workflows strengthens malware analysis, detection, and attribution efforts~\cite{rani2024comprehensive}. Security teams can use it for automated malware triage, where MaLAware processes sandbox reports, extracts key behavioral insights, and generates structured summaries, reducing the manual effort required by analysts. During incident response, MaLAware enhances decision-making by delivering concise insights into a malware’s tactics, such as data exfiltration or persistence mechanisms. This enables teams to respond swiftly by isolating compromised systems or blocking malicious communication channels. Additionally, threat intelligence teams can leverage MaLAware to map observed malware behaviors to known tactics and techniques, enriching security defenses and improving detection rules. In compliance and reporting, MaLAware aids in generating clear, high-level reports for executives, legal teams, and regulatory bodies, ensuring better communication of security threats. 

In practical applications, MaLAware can be deployed in Security Operations Centers (SOC)\footnote{\url{https://csf.tools/reference/nist-sp-800-53/r5/ir/ir-4/ir-4-14/}} to streamline malware analysis, enabling analysts to respond more effectively to threats. 
Managed Security Service Providers (MSSPs) can integrate MaLAware into their services for providing valuable insights to client organizations that lack in-house expertise. Furthermore, threat intelligence platforms can utilize MaLAware to transform raw sandbox reports into actionable intelligence.

\section{MaLAware}
\label{sec:MaLAware}

\begin{figure}[htp]
    \centering
    \includegraphics[width=\columnwidth, height=5.2cm]{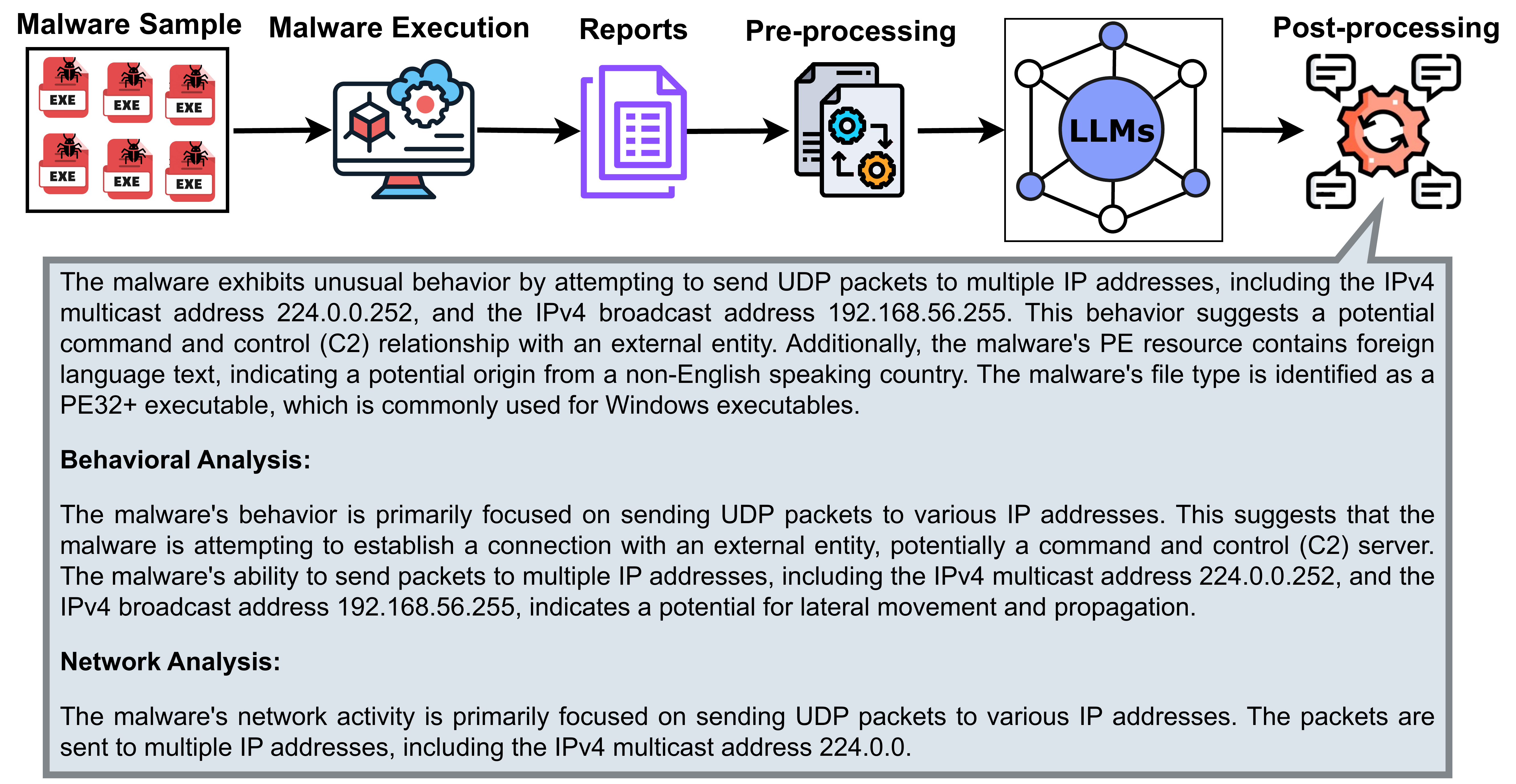}
    \caption{MaLAware Workflow Overview}
    \label{fig:MaLAwareArchitecture}
\end{figure}

MaLAware's architecture incorporates several components to ensure effective operation as shown in Fig.~\ref{fig:MaLAwareArchitecture}. It requires a sandboxing setup to safely execute malware samples in a virtual environment. This is achieved by configuring a cuckoo sandbox on a virtual machine. Various guides and blogs are available to assist with this setup~\cite{cuckoo_docs,hatching_blog,beginning_hacking}.

The cuckoo sandbox operates on an agent-host system. An agent runs on the virtual environment, capturing system changes and operations during malware execution. Simultaneously, another instance on the host machine collects the details shipped by the virtual machine agent. The sandbox generates reports in JSON format, documenting all system changes and recorded operations.

These sandbox reports serve as input for the LLMs. Since the reports include non-relevant details such as timestamps, file sizes, hashes, and analysis duration, a pre-processing script filters out this extraneous information. This reduces the report's text length, optimizing it for the LLMs by minimizing context length and computational complexity.

After pre-processing, the filtered report is passed to the LLM model for analysis. The LLM extracts key events, identifies patterns, and establishes relationships between different actions described in the report. By leveraging its contextual understanding, the model links sequential events, detects dependencies between malicious behaviors, and infers the overall execution flow of the malware. This correlation enables the LLM to reason through the data, distinguishing cause-and-effect relationships and uncovering hidden attack strategies. Finally, it synthesizes these insights into a structured summary, outlining the malware’s actions and their impact during execution.

Finally, the generated summary undergoes a post-processing step to refine its structure, coherence, and completeness. This step ensures appropriate formatting, and logical flow between sentences. The post-processing script produces a structured and readable summary of the malware's malicious actions.

The tool uses 4-bit quantization to reduce memory and computational demands, ideal for resource-constrained environments. It stores model weights in 4-bit precision, uses FP16 for computation, and enables faster inference while preserving performance. This approach lowers resource requirements, boosts energy efficiency, and reduces costs for running large models on limited hardware.

\section{Evaluation}
\label{sec:MaLAwareEvaluation}

\subsection{Dataset}
\label{subsec:MaLAwareGroundTruth}
\begin{figure}
    \centering
    \includegraphics[width=\linewidth, height=4.1cm]{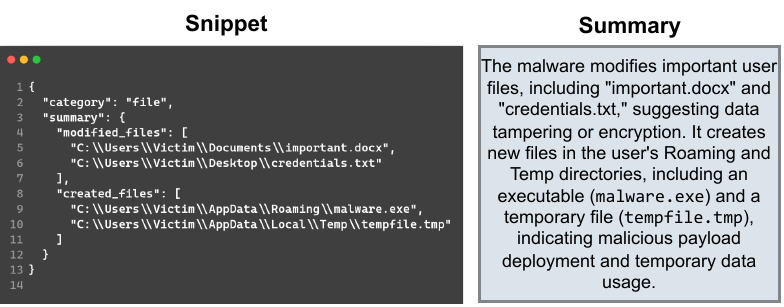}
    \caption{Ground Truth Data Preparation Examples}
    \label{fig:MaLAwaredataGroundtruth}
\end{figure}

We collect malware samples from the public repository MalwareBazaar\footnote{\url{https://bazaar.abuse.ch/}} for this experiment. Samples are selected from diverse families to evaluate LLMs' ability to understand various malicious activities. Preparing the ground truth is challenging, as no existing dataset is publicly available for such a tool. Using our malware expertise, we manually analyze malware behavior. We also leverage online analysis engines like VirusTotal\footnote{\url{https://www.virustotal.com/gui/}} to gain insights and write behavior descriptions. An example of manually narration is shown in Fig.~\ref{fig:MaLAwaredataGroundtruth}. This process results in \nummalware human-written behavior summaries.

\begin{table*}[htbp]
\centering
\begin{threeparttable}
\caption{Performance Comparison of all five implemented LLMs}
\label{tab:MaLAwarePerformanceComparison}
\begin{tabular}{@{}lccccccccccc@{}}
\toprule
\textbf{Model}          & \textbf{R-1}  & \textbf{R-2}  & \textbf{R-L}  & \textbf{BS-P} & \textbf{BS-R} & \textbf{BS-F1} & \textbf{SS}   & \textbf{FKR}   & \textbf{D-1}  & \textbf{D-2}  & \textbf{KM}   \\ \midrule

Qwen2.5-7B-Instruct     & \textbf{0.3876}     & 0.0883     & 0.1893     & \textbf{0.6305}     & \textbf{0.6443}     & \textbf{0.6359}     & \textbf{0.7568}     & 19.1412     & 0.7571     & 0.9503     & 0.2034 \\
Llama-2-7b-hf & 0.1751 & 0.0325 & 0.1147 & 0.5311 & 0.4578 & 0.4906 & 0.5685 & \textbf{40.1621} & 0.6259 & 0.7550 & 0.1167 \\
Llama-3.1-8B-Instruct & 0.3392 & 0.0775 & 0.1787 & 0.6202 & 0.5851 & 0.6016 & 0.7519 & 31.1912 & 0.5690 & 0.7518 & 0.1484 \\
Mistral-7B-Instruct-v0.3 & 0.3648 & \textbf{0.0952} & \textbf{0.1995} & 0.6246 & 0.6040 & 0.6136 & 0.7390 & 35.5210 & \textbf{0.7697} & \textbf{0.9619} & \textbf{0.2333} \\
falcon-7b & 0.1516 & 0.0294 & 0.0959 &  0.5016 & 0.4613 & 0.4754 & 0.5530 & 19.1068 & 0.7560 & 0.8553 & 0.1458 \\

\bottomrule
\end{tabular}

\begin{tablenotes}
\small
\item \textbf{R-1:} Rogue-1, \textbf{R-2:} Rogue-2, \textbf{R-L:} Rogue-L, \textbf{BS-P:} BERTScore Precision, \textbf{BS-R:} BERTScore Recall,
\item \textbf{BS-F1:} BERTScore F1-score, \textbf{SS:} Semantic Similarity, \textbf{FKR:} Flesch-Kincaid Readability, 
\item \textbf{D-1:} Distinct-1, \textbf{D-2:} Distinct-2, \textbf{KM:} Keyphrase Matching
\end{tablenotes}
\end{threeparttable}

\end{table*}

\subsection{Evaluation Metrics}
\label{subsec:MaLAwareEvaluationMetrics}

Comparing generated text with ground truth requires a multi-dimensional evaluation to ensure MaLAware's effectiveness. The evaluation metric selection spans lexical metrics (ROUGE~\cite{lin2004rouge}, BLEU~\cite{papineni2002bleu}), semantic similarity metrics (BERTScore~\cite{zhang2019bertscore}, Semantic Similarity), structural and readability metrics (Flesch-Kincaid Readability~\cite{flesch1948readability}, Distinct-N~\cite{li2016diversity}), and content focus metrics (Keyphrase Matching~\cite{rose2010automatic}).

Lexical metrics assess word and sequence overlap between generated and reference text, with ROUGE evaluating n-gram overlap and focusing on recall to capture key information and maintain alignment. BERTScore measures contextual similarity using pre-trained models like BERT, computing cosine similarity of token embeddings for precision, recall, and F1, ensuring robustness to paraphrasing. Structural and readability metrics evaluate coherence, fluency, and uniqueness, with Distinct-N highlighting n-gram diversity to ensure engaging and meaningful summaries. The content focus metrics verify accurate capture and retention of key information. 

In total, we use \numperformance performance metrics for comprehensive evaluation, detailed below:

\subsubsection{ROUGE-1}
It measures unigram (single word) overlap, providing a basic assessment of word-level content relevance.
\subsubsection{ROUGE-2}
It measures bigram (two consecutive words) overlap, capturing the contextual and sequential alignment between the summaries.
\subsubsection{ROUGE-L}
It evaluates the longest common subsequence (LCS), focusing on the structural similarity and fluency of the text.
\subsubsection{Semantic similarity}
It evaluates the semantic closeness between generated and reference summaries by computing cosine similarity between sentence embeddings generated by models like sentence transformers.

\subsubsection{BERTScore-Precision}
It measures how much of the generated content is relevant to the reference.
\subsubsection{BERTScore-Recall}
It measures how well the generated content captures all key points from the reference.
\subsubsection{BERTScore-F1-Score}
It provides a balanced view of BERTScore-precision and BERTScore-recall.
\subsubsection{Flesch-Kincaid Readability Score}
This metric assesses the readability of a text based on its sentence length and syllable count.  The higher scores indicate easier readability. It ensures the summary is suitable for the target audience.

\subsubsection{Distinct-1}
It is the proportion of unique unigrams (words) in the text.
\subsubsection{Distinct-2}
It is the proportion of unique bigrams (two-word sequences) in the text.
\subsubsection{Keyphrase Matching}
It evaluates key phrase overlap between generated and reference summaries using tools like KeyBERT~\cite{grootendorst2020keybert} to assess core idea alignment.
\\
The metrics discussed above range from 0 to 1, except for the Flesch-Kincaid Readability Score, which ranges from 0 to 100. 
Combination of these metrics provides a holistic evaluation of summarization quality, covering relevance, coherence, fluency, diversity, and readability. This comprehensive approach provides a nuanced understanding of model performance, enabling targeted improvements for better summarization.

\subsection{Results}
\label{subsec:MaLAwareResults}

The evaluation of the models on the prepared dataset using the selected performance metrics is shown in Table~\ref{tab:MaLAwarePerformanceComparison}. The results reveal notable differences in the models' ability to generate effective explanations and summaries.

Qwen2.5-7B-Instruct stands out as the top-performing model. It achieves the highest scores in ROUGE-1 (0.3876), BERTScore Precision (0.6305), and BERTScore F1 (0.6359), demonstrating its ability to generate coherent and contextually accurate text. The model excels in Semantic Similarity (0.7568), showing its strength in producing text that aligns closely with the reference. These results highlight Qwen2.5-7B-Instruct as a strong model for tasks requiring high-quality, semantically relevant content.

Mistral-7B-Instruct-v0.3 follows closely, with notable performance in ROUGE-2 (0.0952) and ROUGE-L (0.1995). The model's strength lies in its readability and ability to produce diverse summaries, reflected in its Keyphrase Matching score of 0.2333 and Distance-2 of 0.7697. These scores indicate that Mistral-7B-Instruct-v0.3 generates summaries that are not only accurate but also easy to read and understand. Despite these strengths, it does not outperform Qwen2.5-7B-Instruct in terms of semantic alignment and fluency.

Llama-3.1-8B-Instruct and Llama-2-7b-hf show weaker performance compared to Qwen2.5-7B-Instruct and Mistral-7B-Instruct-v0.3. Llama-3.1-8B-Instruct performs moderately in ROUGE-1 (0.3392) and ROUGE-2 (0.0775), and its Semantic Similarity (0.7519) is lower than that of the top models, indicating that it struggles to maintain contextual relevance in its outputs. While it provides semantically precise content, it lags in diversity and coherence, making it less versatile for broad applications. Llama-2-7b-hf performs the worst across most metrics, especially in ROUGE and BERTScore, suggesting its limited capability in producing meaningful and contextually aligned summaries.

Falcon-7B performs competitively in D-1 and D-2, with scores of 0.7560 and 0.8553, respectively. This shows its potential for distinctive words in the summaries. However, its weak scores in ROUGE (e.g., R-1: 0.1516) and BERTScore reflect its inability to generate summaries that closely match reference text. This highlights the difficulty of balancing readability with semantic richness, as Falcon-7b struggles to generate detailed, contextually relevant explanations.

In conclusion, Qwen2.5-7B-Instruct is the best-performing model and demonstrates superior semantic and contextual accuracy and fluency. Mistral-7B-Instruct-v0.3 excels in generating readable and diverse summaries but falls short in overall alignment. Llama-3.1-8B-Instruct and Llama-2-7b-hf show limitations in semantic similarity and diversity, while Falcon-7b faces challenges in balancing readability with meaningful content generation. These findings highlight the importance of selecting models based on task-specific requirements, whether it be semantic accuracy, readability, or content diversity.

\section{Limitation \& Future Work}
\label{sec:MaLAwareLimitationFutureWork}

The current version of the presented tool has some limitations that present opportunities for future improvement in subsequent versions. A significant challenge lies in the high cost and resource demands of working with LLMs, making evaluation time and resource intensive. This complexity becomes even more pronounced when scaling to larger datasets or deploying models in real-time environments. Enhancing computational efficiency is crucial to reduce overhead and improve the model's practicality for real-world applications.

The reliance on open-source models is another limitation. While black-box alternatives like GPT-4, GPT-4o, and Claude often deliver better performance, they limit transparency and customization. These models also come with added costs for services. Balancing performance benefits with financial constraints is critical when evaluating their suitability. Currently, we focus solely on open-source models. In future versions, we plan to integrate black-box options, allowing users to choose models based on their preferences.

Another limitation is the lack of fine-tuning for this specific downstream task. The evaluated models were pre-trained for general purposes and  not tailored for malware explanation, show inconsistent performance due to a lack of domain-specific understanding. Our future efforts will focus on task-specific fine-tuning to enhance the precision, diversity, and coherence of the generated explanations.

For the future, we focus on task-specific fine-tuning, efficient architectures, and integrating black-box models. We also aim to expand the human-written ground truth dataset to evaluate future versions on a larger scale, increasing confidence in MaLAware's effectiveness. These steps enhance the models' explanatory power, efficiency, and adaptability for diverse malware analysis scenarios. This approach ensures a robust and reliable malware explanation system tailored to various application needs.

\section{Related Work}
\label{sec:MaLAwareRelatedWork}

Existing research on malware analysis often utilizes AI and language models for tasks like malware detection or family classification. Language model-based approaches primarily focus on code-level analysis, dynamic API sequences, or vulnerability identification. However, their outputs frequently lack clarity, making it difficult for a wider audience to grasp the malware's actions and intent.

Richards~\cite{richards2023} explores LLMs for reverse engineering malware, showcasing their ability to interpret decompiled code. While effective, this approach produces technical outputs that lack user-friendly explanations for non-technical audiences. Yan et al.~\cite{yan2023prompt} generate API-level explanations and feed them into a BERT classifier for malware detection. While effective, these methods primarily cater to experts for malware detection and fail to provide behavior insights in human-readable text. Our method addresses these gaps by using LLMs to generate clear, concise narratives that explain malware behavior in human-understandable terms. This enables cybersecurity professionals and non-experts alike to understand and respond to threats effectively.

Pordanesh and Tan~\cite{pordanesh2024exploring} highlight the limitations of LLMs, such as GPT-4, in producing detailed analyses for binary reverse engineering. Our approach moves beyond static outputs by integrating context-aware language generation. This shift focuses on behavioral insights, making our method more practical for real-world cybersecurity applications.

Our work advances the understanding of malware behavior by providing explanations that are actionable, accessible, and contextually accurate. This approach empowers users with varying technical expertise to make informed decisions in addressing cyber threats. By bridging the gaps in existing research, our method marks a significant step forward in leveraging LLMs for cybersecurity applications.

\section{Conclusion}
\label{sec:MaLAwareConclusion}
In this study, we present a LLM-powered framework to automatically comprehend the malignant action taken by malicious software in terms of their behaviour. We build a tool name MaLAware based on the presented framework. We also evaluate the effectiveness of state-of-the-art language models in generating explanations for malware behaviors, emphasizing the importance of contextual accuracy, readability, and diversity. Our findings show that Qwen2.5-7B-Instruct performs best, excelling in ROUGE and BERTScore metrics. It generates fluent, relevant, and semantically accurate summaries. Mistral-7B-Instruct-v0.3 follows closely, performing well in readability and diversity.
This work underscores the growing potential of leveraging advanced language models to enhance understanding in cybersecurity, particularly in aiding analysts to understand and interpret complex malware behavior. 
In future work, we aim to enhance malware analysis through fine-tuning, efficient architectures, black-box model integration, and expanded datasets. These efforts will address existing gaps, enabling the development of systems that deliver actionable insights and strengthen cybersecurity resilience.



\end{document}